\documentclass[prb,aps]{revtex4}
\usepackage{graphicx}
\begin{document}

\title{Cluster approach study of intersite electron correlations in pyrochlore and checkerboard
lattices}

\author{V. Yushankhai}
\affiliation{Max-Planck-Institut f\"ur Physik Komplexer Systeme,
D-01187 Dresden, Germany}
\affiliation{Joint Institute for Nuclear
Research, 141980 Dubna, Russia}

\author{P. Fulde}
\affiliation{Max-Planck-Institut f\"ur Physik Komplexer Systeme,
D-01187 Dresden, Germany}

\author{P. Thalmeier}
\affiliation{Max-Planck-Institut f\"ur Chemische Physik fester
Stoffe, D-01187 Dresden, Germany}

\date{\today}

\begin{abstract}
To treat effects of electron correlations in geometrically
frustrated pyrochlore and checkerboard lattices, an extended single-orbital
Hubbard model with nearest neighbor hopping $\sim t$ and
Coulomb repulsion $\sim V$ is applied. Infinite on-site
repulsion, $U\to\infty$, is assumed, thus  double  occupancies of sites
are forbidden completely in the present study. A variational
Gutzwiller type approach is extended to examine correlations due
to short-range $V-$interaction and a cluster approximation is
developed to evaluate a variational ground state energy of the
 system. Obtained analytically in a special case of quarter band
filling appropriate to LiV$_2$O$_4$,  the resulting simple expression
describes the ground state energy in the regime of intermediate and
strong  coupling  $V$. Like in the Brinkman-Rice theory
based on the standard Gutzwiller approach to the Hubbard model,
the mean value of the kinetic energy is shown to be reduced
strongly as the coupling $V$ approaches a critical value $V_{c}$.
This finding may contribute to explaining the observed heavy
fermion behavior in LiV$_2$O$_4$.
\end{abstract}


\maketitle

\section{Introduction}
The pyrochlore lattice is a three-dimensional network of
corner-sharing tetrahedra formed by B-cations in pyrochlore
oxides A$_2$B$_2$O$_7$, in cubic spinel oxides AB$_2$O$_4$ and in cubic
Laves-phase intermetallic compounds AB$_2$. These compounds exhibit
a wide variety of physical properties ranging from magnetic
insulator through bad metal to superconductor with relatively high
transition temperature. One of the most spectacular properties in
this family of $d$-systems is the heavy fermion (HF) behavior found
in LiV$_2$O$_4$ and Y(Sc)Mn$_2$ at low temperatures.
\cite{Kondo97,Wada87,Fisher92} To understand the nature of HF
quasiparticles in Y(Sc)Mn$_2$, it was suggested
\cite{Isoda00,Fujimoto01} that effects of geometrical frustration
on the itinerant electron system in the pyrochlore lattice are of
crucial importance.

In fact, in the pyrochlore lattice with a single orbital on each
site, a special lattice geometry manifests itself through a flat
band on the top of the dispersive one-particle electron
spectrum.\cite{Isoda00} In the half-filled case, the Fermi level
touches the flat band singularity from below. In actual band
structure calculations of a multi-orbital $d$-system, this
singularity is removed  and, instead, a sharply peaked feature in
the density of states is seen. By treating the problem of
Y(Sc)Mn$_2$ within a single band pyrochlore Hubbard model in a
weak correlation limit and assuming proximity of the Fermi level
to the peaked feature in the density of states, the authors
\cite{Isoda00,Fujimoto01} suggested a new scenario for the HF-like
behavior of the frustrated system. In contrast, in LiV$_2$O$_4$ 
the mean  occupancy of the $V$-ion $d$-shell equals 1.5 and corresponds to
the quarter-filling of the almost three-fold degenerate $t_{2g}$
band.\cite{Singh99} Therefore, in  this compound the Fermi level
is pushed down strongly from  the 'flat-band' peculiarity and the
above arguments for the HF quasiparticle formation are not
applicable any more. In this respect, even by taking into account
strong on-site Coulomb correlation effects, not included in the
band structure calculations, one cannot cure the situation.

Several scenarios for explaining HF properties in LiV$_2$O$_4$,
including the one based on multi-component
fluctuations\cite{Yamashita03},  are
proposed in the literature and reviewed briefly in Ref.\onlinecite{Fulde03}.
While the strong  on-site Coulomb interaction $\sim U$ is
incorporated necessarily in  most of these scenarios, effects of
inter-site Coulomb correlations on the HF properties in the
pyrochlore lattice have been studied much less. The latter effects
in the form of Anderson's 'tetrahedron rule' \cite{Anderson56} are
used explicitly in Ref.\onlinecite{Fulde01} to build a physical picture of
low temperature properties in LiV$_2$O$_4$. According to this
picture, Coulomb repulsions of electrons on nearest neighbor
sites suppress charge fluctuations appreciably and a state of
slowly fluctuating and nearly decoupled chain-like and ring
configurations of spins $s=1/2$ is formed to give a large low
temperature spin entropy. However, the mechanism for suppressing
charge fluctuations in the pyrochlore lattice has not been
elaborated  in detail.

In the present paper, we address  this problem and
concentrate on the study of charge degrees of freedom and their
possible role in forming heavy quasiparticles. For this purpose,
the potentially important role of inter-site electron Coulomb
interaction in the pyrochlore lattice  is emphasized in a
particular case of quarter band filling appropriate to
LiV$_2$O$_4$. We use an extended single-orbital Hubbard model
composed of the kinetic term with a tight-binding nearest neighbor
(n.n.) hopping amplitude $t$, strong Hubbard term $\sim U$ and
Coulomb term describing n.n. electron repulsion $\sim V$ of
variable strength with respect to $t$. In spite of its serious
simplification, this minimal model is believed to capture generic
effects of strong inter-site Coulomb correlations inherent to more
realistic pyrochlore electronic models including several
$d$-orbitals on each lattice site. Noting that $U$ is the largest
energy parameter, we simply put $U\rightarrow \infty$ in our
calculations. At $U\rightarrow \infty$ and far away from
half-filling, one expects the system to be in a metallic state at
weak coupling $V$. Our aim is to show that in the strong coupling
limit,  $V\gg t$, an effective electron bandwidth in the
quarter-filled system can be strongly reduced due to short-range
charge correlations. This regime of a  strongly correlated metal
can be understood as a state of almost localized electrons.

Short-range spin correlations are neglected here because the limit
$U\rightarrow \infty$ is assumed. Instead, we concentrate completely
on the problem of
short-range \textit{charge} correlations which are believed to
provide the dominant mechanism for strong suppression of the
electron kinetic motion in the pyrochlore lattice and, therefore,
may contribute to the heavy quasiparticle formation.

 To conclude this Section, we outline briefly the
 structure of the paper and the method used. We extend
 the Gutzwiller variational approach to a form that allows us to
 treat the inter-site correlation problem. For completeness, we
 involve in our consideration also the 2D checkerboard lattice
 which is a planar analogue of the 3D pyrochlore lattice. Our
 method is applicable equally to both cases and relies on a cluster
 character of the pyrochlore (checkerboard) lattice structure,
 which means that each tetrahedron (plaquette) is regarded as an
 elementary entity. In Section II, the underlying electronic
 model is defined and rewritten in a cluster representation. The
 ground state variational wave function of the interacting system
 is constructed  by applying to the system of
 noninteracting electrons two kinds of projectors. The first is the
 Gutzwiller projector taken in the limit $U\rightarrow\infty$ to
 eliminate double occupancy of sites. The second projector is defined
 in terms of cluster occupancy operators in such a way that the
 weights of different charge configurations in each cluster are
 controlled by a correlation strength parameter $\eta$ which is the
 only variational parameter in our theory.

In Section III, a variational ground state energy functional is
defined and a factorization procedure of its approximate
calculation is discussed in detail. Within this  context, the
concept of the basic cluster in the pyrochlore (checkerboard)
lattice is introduced and a mechanism for suppressing of
inter-cluster charge fluctuations is explained. In Section IV, the
calculations are performed and the resulting ground state energy
as a function of the coupling $V$ is analyzed. This  analysis
shows that the mean kinetic energy in the system is
strongly renormalized and tends to zero as $V$ approaches some
critical value $V_{c}$.
Also, our variational theory suggests,
that electron localization transition occurs at $V=V_c$, and for
$V>V_c$ the electrons are in a fully localized insulating state for which
quantum charge fluctuations are completely missed. The results are
discussed in Section V. By drawing a close
analogy with the Brinkman-Rice localization transition predicted in
the variational study\cite{Brinkman70}  of  the standard
Hubbard model, we argue that improper description of
the actual metal-insulator transition does not invalidate our cluster
variational theory in all respects. In particular, the theory is still valid
in describing a metallic strongly correlated (the not yet localized)
regime for $V\lesssim V_c$. Concluding remarks can be also found in
Section V.

\section{Model formulation and trial wave function}
Let us start with an extended single-orbital Hubbard model in the
pyrochlore or checkerboard lattice
\begin{eqnarray}
H=-t \sum_{<ij>,\sigma}\left(c^\dagger_{i\sigma}c_{j\sigma} +
c^\dagger_{j\sigma}c_{i\sigma}\right) + V\sum_{<ij>}n_in_j  +
U\sum_{i} n_{i\uparrow}n_{i\downarrow}, \label{a1}
\end{eqnarray}
where the summation in the first two terms is taken over the n.n.
sites, each pair $<ij>$ is counted once;
$c^\dagger_{i\sigma}(c_{j\sigma})$ is the creation (annihilation)
electron operator with spin $\sigma=\uparrow,\downarrow$ and
$n_i=\sum_\sigma c^\dagger_{i\sigma}c_{i\sigma}$. We are
interested in the model regime far away from the band half-filling
in the limit $U\to\infty$.

 Let us consider a plaquette in the checkerboard or a tetrahedron in the
pyrochlore  lattices as an \textit{elementary} $\mathcal{K}$-th
cluster, each contains four sites numbered as $k=1,...,4$. Each
lattice site belongs to two clusters and thus the
neighboring clusters overlap. If $L$ is the
number of the lattice sites, then $L/2$ is the number of
overlapping clusters, i.e. ${\mathcal{K}}=1,2,\ldots,L/2$. Now, the
first two terms in the Hamiltonian (\ref{a1}) can be rewritten in
the cluster representation
\begin{eqnarray}
&&H_{t-V}=\sum_{{\mathcal{K}}=1}^{L/2}H_{\mathcal{K}}, \qquad
H_{\mathcal{K}}= T_{\mathcal{K}} + V_{\mathcal{K}},\nonumber\\
&&T_{\mathcal{K}}=-t\sum_{\stackrel{k\neq k'}{(k,k')\in
{\mathcal{K}}}} \sum_\sigma
c^\dagger_{k\sigma}c_{k'\sigma}, \label{a2}\\
&&V_{\mathcal{K}} = \frac{V}{2}\sum_{\stackrel{k\neq k'}{(k,k')\in
{\mathcal{K}}}}n_k n_{k'}, \qquad n_k=\sum_\sigma
c^\dagger_{k\sigma}c_{k\sigma}.\nonumber
\end{eqnarray}
Since $U\to\infty$, the double  occupancy of sites is forbidden and the
unit operator on each $k$-th site is
\begin{eqnarray}
&&1_k=P_k(0) + P_k(1),\nonumber\\
&&P_k(0)=(1-n_{k\uparrow})(1-n_{k\downarrow}), \label{a3}\\
&&P_k(1)=\sum_{\sigma =\uparrow,\downarrow}P_{k\sigma}(1), \qquad
P_{k\sigma}=n_{k\sigma}(1-n_{k\bar\sigma}),\nonumber
\end{eqnarray}
where $P_k(0)$ and $P_k(1)$ are the projection operators onto the
empty and singly occupied $k$-th site states, respectively;
$\bar\sigma = -\sigma$. Further, to distinguish cluster states
with occupancies $n=0,\ldots,4$, we define the cluster projection
operators
\begin{eqnarray}
&&{\mathcal P}_{\mathcal{K}}(0)=\prod_{k=1}^{4}{\mathrm P}_{k}(0),\nonumber\\
&&{\mathcal P}_{\mathcal{K}}(1) =\sum_{k=1}^{4}{\mathrm{P}}_{k}(1)
\left[\prod_{k'\neq k}{\mathrm P}_{k'}(0)\right],\nonumber\\
&&{\mathcal P}_{\mathcal{K}}(2)=
\sum_{k<k'}{\mathrm{P}}_{k}(1){\mathrm{P}}_{k'}(1)
\left[\prod_{k''\neq k,k'}{\mathrm P}_{k''}(0)\right],\label{a4}\\
&&{\mathcal P}_{\mathcal{K}}(3) =\sum_{k=1}^{4}{\mathrm{P}}_{k}(0)
\left[\prod_{k'\neq k}{\mathrm P}_{k'}(1)\right],\nonumber\\
&&{\mathcal P}_{\mathcal{K}}(4)=\prod_{k=1}^{4}{\mathrm P}_{k}(1),
 \nonumber
\end{eqnarray}
where ${(k,k',k'')\in {\mathcal{K}}}$ is implied. Disregarding the
spins, we note that the configurational space of a cluster
contains in total 16 allowed charge configurations. For instance,
the  occupancy state $n=2$ is composed of 6 configurations.
Taking into account the frustration of the system, we suggest that
the basic operators (\ref{a4}) and the associated cluster
occupancies in the lattice suffice to capture the correlated
metallic regime of the model, as suggested by the form (\ref{a6})
of $V_{\mathcal{K}}$ term below.

In terms of the projection operators (\ref{a4}), the kinetic and
the Coulomb terms in (\ref{a2}) can be presented as follows
\begin{equation}\label{a5}
T_{\mathcal{K}}=\sum_{n=1}^3T_{\mathcal{K}}(n) =
\sum_{n=1}^3{\mathcal{P}}_{\mathcal{K}}(n)T_{\mathcal{K}}{\mathcal{P}}_{\mathcal{K}}(n),
\end{equation}
\begin{equation}\label{a6}
  V_{\mathcal{K}}=V\left[{\mathcal{P}}_{\mathcal{K}}(2) +
  3{\mathcal{P}}_{\mathcal{K}}(3)+
  6{\mathcal{P}}_{\mathcal{K}}(4)\right],
\end{equation}
where the diagonal form of $T_{\mathcal{K}}$ occurs because any
electron hopping within a cluster does not change its charge
occupancy. Finally, we define the cluster number operator
$N_{\mathcal{K}}=\sum_{k\in{\mathcal{K}}}n_k$, and note that
the mean value $\langle N_{\mathcal{K}}\rangle =2$ corresponds to
the quarter band filling.

A trial ground state wave function of the interacting system is
constructed below in two steps. First, in the limit $V=0$, we
introduce the Gutzwiller projected  Fermi sea $(U\to\infty)$:
\begin{equation}\label{a10}
|\psi\rangle_{V=0} = \frac{1}{\sqrt{\Omega}} \prod_{k=1}^L\left[1
- n_{k\uparrow}n_{k\downarrow}\right]|\Phi_0\rangle
\equiv|\psi_0\rangle,
\end{equation}
where $|\Phi_0\rangle$ is a single Slater determinant describing a
metallic state for $N=L/2$ noninteracting electrons; $\Omega$ is
the norm. Since the system is away from half-filling, the
Gutzwiller projectors \cite{Gutzwiller65,Vollhardt84} in
(\ref{a10}) are expected to lead to a moderate renormalization of  metallic
properties. A regime of strongly correlated metal at quarter band
filling may occur in an interacting system with strong inter-site
Coulomb repulsion $V$.

To take into consideration  electron correlations due to $V\neq 0$, we
define below a product of generalized Gutzwiller projectors
$G_{\mathcal{K}}(\eta)$
acting in the occupancy space of the $\mathcal{K}$-th clusters
(${\mathcal{K}}=1,2,\ldots,L/2$);  here
$\eta$ is the variational parameter varying in the range $0 \leq \eta
\leq 1$.
To justify the proper choice of $G_{\mathcal{K}}(\eta)$, let us start
with the limit of extreme coupling, $V/t \to\infty$. In this limit,
the ground state cluster charge configurations are those obeying the
'tetrahedron rule', i.e. each cluster is occupied strictly with two
electrons, and the ground state energy is $V$ per cluster.  Therefore,
 $G_{\mathcal{K}}(\eta) = {\mathcal
 P}_{\mathcal{K}}(2)$ and our convention is that $\eta = 0$ in this
limit. At finite $t$, electron hopping processes  create exited charge
configurations of higher, $n=3,4$ and lower, $n=0,1$, cluster
occupancies.  To ascribe optimal weights to the exited configurations,
we use the following prescription. Consider first a pair of clusters,
 ${\mathcal{K}}_1$ and ${\mathcal{K}}_2$, occupied with $n_1=1$ and
 $n_2=3$ electrons. By noting
 that the potential energy of such a fluctuation is  $3V$, which
 exceeds the energy $2V$ of the dominant configuration of two
 electrons per cluster by $V$, we ascribe to it a weight factor $\eta
 \leq 1$. The cost of the potential energy of the more extreme charge
 fluctuation of a pair of clusters, i.e. $n_1=0$ and $n_2=4$, is $4V$,
 and therefore we associate the weight factor ${\eta}^4$ with this
 fluctuation.  More complex many-cluster excited charge
 configurations  can be
 easily recognized to be combinations of the elementary ones discussed
 above. These observations are comprised in the following form of a
 generalized cluster  Gutzwiller projector
\begin{equation}\label{a43}
G_{\mathcal{K}}(\eta)= {\eta}^2{\mathcal{P}}_{\mathcal{K}}(0) +
\sqrt{\eta}{\mathcal{P}}_{\mathcal{K}}(1) + {\mathcal{P}}_{\mathcal{K}}(2) +
\sqrt{\eta}{\mathcal{P}}_{\mathcal{K}}(3) + {\eta}^2{\mathcal{P}}_{\mathcal{K}} (4),
\end{equation}
and of a trial ground state wave function  defined as
\begin{equation}\label{a18}
|\psi(\eta)\rangle =
\prod_{\mathcal{K}}^{L/2}G_{\mathcal{K}}(\eta)|\psi_0\rangle
=|\psi\rangle,
\end{equation}
where $0\leq \eta \leq 1$. If $\eta=1$, the projector
$G_{\mathcal{K}}(\eta)$ becomes the  unity operator, which corresponds to
the limit of zero coupling, $V=0$.

Note, the form (\ref{a43}) of a generalized Gutzwiller projector
and, hence, the trial wave function (\ref{a18}) are inspired by
the form (\ref{a6}) of $V_{\mathcal{K}}$ term and Anderson's
reasoning  \cite{Anderson56}  on a macroscopical degeneracy of the ground state charge
configurations in the limit of the extreme coupling, $V/t\to\infty$. We
expect the wave function ansatz (\ref{a18})  to be valid
down to intermediate coupling values $V \sim zt$, where $z=6$ is
the pyrochlore (checkerboard) lattice coordination number.
It is much less appropriate in the limit of a weak coupling $V
\lesssim t$, where the kinetic energy effects become dominant and the
arguments based on the potential energy counting only are not sufficient.

\section{Variational ground state energy and factorization procedure}
With definitions (\ref{a1}) and (\ref{a2}), the ground state
energy of the lattice system ${\mathcal{E}}(\eta)=\langle\psi|H|
\psi\rangle / \langle\psi|\psi\rangle$
can be written as a sum of cluster energies
\begin{equation}\label{a20}
{\mathcal{E}}(\eta)=\frac{L}{2}E(\eta);\qquad
E(\eta)=\frac{\langle\psi(\eta)|H_{\mathcal{K}}|\psi(\eta)\rangle}
{\langle\psi(\eta)|\psi(\eta)\rangle},
\end{equation}
where the translational invariance of the wave
function  (\ref{a18}), i.e. $\langle\psi|H_{\mathcal{K}}|\psi\rangle=
\langle\psi|H_{\mathcal{K}'}|\psi\rangle$
for ${\mathcal{K}}\neq{\mathcal{K}'}$, is taken into account.
From now on, the expectation energy value calculation reduces to a
local problem.

To evaluate $E(\eta)$, an approximate procedure is
developed and applied below . The starting point is to divide the
lattice into two parts. The smaller one, which is refereed below
to as the \textit{basic cluster} ($bc$), contains five connected
elementary clusters, as explained in Fig.1 for the checkerboard
lattice. According to this picture, the same hopping parameter $t$ and
Coulomb interaction constant $V$ should be ascribed to the horizontal,
vertical and diagonal bonds (linear segments) connecting neighboring
lattice sites.
One can easily realize that the basic cluster in Fig.1
has its 3D counterpart formed by five corner-sharing tetrahedra in
the pyrochlore lattice, the former being a planar projection of
the latter. Therefore, both cases can be treated on an equal
footing. With this division, we suggest that the  essential
effects of the local interactions involved in $E(\eta)$ are
captured by a basic cluster.

Below, a factorization procedure for
the trial wave function (\ref{a18}) is used, which reduces the
calculation of expectation value $E(\eta)$  from the lattice to
a basic cluster.
In the present study, we take advantage of another simplifying
assumption which amounts to the neglect of the  spatial charge
correlations in the \textit{starting} wave function (\ref{a10}).
Such an approximation is inherent also to the
Gutzwiller approach study of standard Hubbard model. Actually, as
shown in Ref.\onlinecite{Ogata75}, when calculating the hopping matrix
elements (the interacting $U$-term is accounted for exactly) in the
$N$-particle state $|\Phi_0\rangle$ of noninteracting electrons,
the neglect of the configuration dependence at this stage is
equivalent to all assumptions made by Gutzwiller
\cite{Gutzwiller65} and leads to an identical
result.\cite{Ogata75} The approximation is controlled by the
parameter $1/z$, where $z$ is the lattice coordination number; it
gives an exact result in the limit of infinite spatial dimension
and requires only very small corrections for three-dimensional
lattices.\cite{Vollhardt84,Metzner88} Though the wave function
(\ref{a10}) describes an electron system with extremely strong
on-site repulsion, we expect that foregoing arguments are still
valid provided the electron concentration is sufficiently far away
from the half filled band case.

After explaining the very essence of the approximation chosen to
calculate the expectation energy value $E(\eta)$, we
proceed now with its evaluation. First, a basic cluster is
selected as a lattice fragment composed of a central elementary
$0$-cluster whose four sites are shared with four (I,...,IV)
side elementary clusters, as depicted in Fig.1. From now
on, the trial state (\ref{a18}) can be written in the following
form
\begin{equation}\label{a21}
|\psi(\eta)\rangle =
G_{{\mathcal{K}}_0}(\eta)G_{\left\{{\mathcal{K}}\right\}}
(\eta)G_{\left\{{\mathcal{K}'}\right\}}(\eta)|\psi_0\rangle,
\end{equation}
where $G_{{\mathcal{K}}_0}$ is referred to the central $0$-cluster,
$G_{\left\{{\mathcal{K}}\right\}}=\prod_{{\mathcal{K}}=I}^{IV}G_{\mathcal{K}}$
is for four side clusters and
$G_{\left\{{\mathcal{K}'}\right\}}=\prod_{{\mathcal{K}'}(\neq{{\mathcal{K}}_0,
{\mathcal{K}}})} G_{\mathcal{K}'}$
is for the remaining system. The factorization procedure now
reads as follows
\begin{eqnarray}
E(\eta)&\simeq&\frac{\langle\psi_0|G_{\left\{{\mathcal{K}}\right\}}G_{{\mathcal{K}}_0}
H_{{\mathcal{K}}_0}G_{{\mathcal{K}}_0}G_{\left\{{\mathcal{K}}\right\}}|\psi_0\rangle\langle\psi_0|
G^2_{\left\{{\mathcal{K}'}\right\}}|\psi_0\rangle}{\langle\psi_0|
G^2_{\left\{{\mathcal{K}}\right\}}G^2_{\left\{{{\mathcal{K}}_0}\right\}}|\psi_0\rangle
\langle\psi_0|
G^2_{\left\{{\mathcal{K}'}\right\}}|\psi_0\rangle}=\nonumber\\
&=&\frac{\langle\psi_0|G_{\left\{{\mathcal{K}}\right\}}G_{{\mathcal{K}}_0}
H_{{\mathcal{K}}_0}G_{{\mathcal{K}}_0}G_{\left\{{\mathcal{K}}\right\}}|\psi_0\rangle}{\langle\psi_0|
G^2_{\left\{{\mathcal{K}}\right\}}G^2_{\left\{{{\mathcal{K}}_0}\right\}}|\psi_0\rangle}=
\frac{\langle\psi_{bc}|H_{{\mathcal{K}}_0}|\psi_{bc}\rangle}{\langle\psi_{bc}|\psi_{bc}\rangle},
\label{a22}
\end{eqnarray}
where the trial state $|\psi_{bc}\rangle$ in the cluster
approximation is taken in the form
\begin{equation}\label{a23}
|\psi_{bc}(\eta)\rangle=G_{{\mathcal{K}}_0}(\eta)
G_{\left\{{\mathcal{K}}\right\}}(\eta)|\psi_0\rangle,
\end{equation}
and the labelling {\it bc} is due to a distinguished basic cluster
either in the pyrochlore or in the checkerboard lattice. Both,
three- and two-dimensional basic clusters have the same lattice
connectivity and, therefore, in our study are treated equally. We
believe that the basic cluster chosen is inevitably a minimal one
for the present problem. This means that it could not be reduced
further, for instance, to the size of  4-site elementary cluster,
otherwise the most important correlation effect of the kinetic
energy reduction would be lost. Actually, when calculating
$\langle\psi_{bc}|T_{{\mathcal{K}}_0}|\psi_{bc}\rangle/
\langle\psi_{bc}|\psi_{bc}\rangle/ = \langle
T_{{{\mathcal{K}}_0}}\rangle_{bc}$, one finds that the electron
hopping processes involved in $T_{{\mathcal{K}}_0}$ do not change
an occupancy of the ${\mathcal{K}}_0$ -th cluster, but any of them
changes simultaneously  occupation numbers  of a pair of side
clusters. As a result, the side cluster charges deviate from their
mean value $n=2$.  With increasing coupling constant $V$, however,
the cluster states of high, $n=3,4$ and low, $n=0,1$, occupancies
are suppressed, and thus the hopping processes within the
${\mathcal{K}}_0$-th cluster are hampered. As a consequence, the
expectation value of the kinetic term $\langle
T_{{{\mathcal{K}}_0}}\rangle_{bc}$ is reduced strongly in a way
similar to that  found in the standard Hubbard model treated
within the Gutzwiller approximation. In the latter case, a minimal
basic cluster is  2-site one as suggested by Razafimandimby.
\cite{Raz82}

In order to  facilitate evaluation of
$E(\eta)$, we resort to a new diagonal form of the operator
product $G_{{\mathcal{K}}_0}G_{\{{\mathcal{K}}\}}$ with respect to
the occupancy number $n$ of the central cluster ${\mathcal{K}}_0$.
For this purpose, let us consider side clusters
${\mathcal{K}}=\mathrm{I},\ldots,\mathrm{IV} $ and their
labels  $\mathcal{K}$ as  composite ones,
${\mathcal{K}}=(k,\overline{k})$, for instance,
$\mathrm{I}=(1,\overline{1})$, etc. Here $k$ denotes a site common
both to the $\mathcal{K}$-th and the central ${\mathcal{K}}_0$
clusters, while
$\overline{k}$  refers to the complementary 3-site fragment of
the ${\mathcal{K}}-$th side  cluster, as shown in Fig.~\ 1. For the
$\overline{k}-$th fragment, let $|\overline{k},m\rangle$ be a
quantum state with an electron occupancy $m(=0,..,3)$ and
${\mathcal{P}}_{\overline{k}}(m)=|\overline{k},m\rangle\langle\overline{k},m|$
are the corresponding projection operators. In terms of
${\mathcal{P}}_{\overline{k}}(m)$, the
projection operators ${\mathcal{P}_{\mathcal{K}}}(n)$ of  side
$\mathcal{K}$-clusters can be now written in a split form
${\mathcal{P}_{\mathcal{K}}}(n)=P_{k}(0){\mathcal{P}}_{\overline{k}}(n)
+P_{k}(1){\mathcal{P}}_{\overline{k}}(n-1)$,
with the convention
${\mathcal{P}}_{\overline{k}}(-1)={\mathcal{P}}_{\overline{k}}(4)=0$
assumed. With this redefinition of
${\mathcal{P}}_{\mathcal{K}}(n)$, the projectors $G_{\mathcal{K}}$
of side clusters are also split  in two terms, each selecting one of two
possible occupancies, $n=0,1$, of the $k-$th site common to a
given $\mathcal{K}$ and the central ${\mathcal{K}}_0$ clusters:
\begin{equation}\label{a26}
G_{\mathcal{K}}(\eta)=P_{k}(0)Q_{\overline{k}}(\eta)
+ P_k(1)R_{\overline{k}}(\eta).
\end{equation}
Here $Q_{\overline{k}}$ and $R_{\overline{k}}$ operate in the
charge configuration space of the $\overline{k}-$th fragment
\begin{eqnarray}
Q_{\overline{k}}(\eta)&=&{\eta}^2{\mathcal{P}}_{\overline{k}}(0) +
\sqrt{\eta}{\mathcal{P}}_{\overline{k}}(1) +
 {\mathcal{P}}_{\overline{k}}(2) +
\sqrt{\eta} {\mathcal{P}}_{\overline{k}}(3),\nonumber\\
R_{\overline{k}}(\eta)&=&\sqrt{\eta} {\mathcal{P}}_{\overline{k}}(0) +
{\mathcal{P}}_{\overline{k}}(1) +
\sqrt{\eta} {\mathcal{P}}_{\overline{k}}(2) +
{\eta}^2 {\mathcal{P}}_{\overline{k}}(3).\label{a27}
\end{eqnarray}
with the following obvious  properties,
$ Q^2_{\overline{k}}(\eta)=Q_{\overline{k}}(\eta^2)$ and
$R^2_{\overline{k}}(\eta)=R_{\overline{k}}(\eta^2)$.

It is not a difficult task to rewrite the operators
${\mathcal{P}}_{\overline{k}}(m)$ and their linear combinations
(\ref{a27}) through the site projection operators (\ref{a3}),
which is useful for further calculations.  The weight factors in
(\ref{a27}) serve as a measure of the correlation strength in
dependence on the $\overline{k}-$th fragment charge occupancy.
Finally, we use the definition (\ref{a43}) for
$G_{{\mathcal{K}}_0}$ and the presentation (\ref{a26}) for $G_{\mathcal{K}}$ to
write down the following form of the product
$G_{{\mathcal{K}}_0}G_{\{{\mathcal{K}}\}}$:
\begin{equation}\label{a29}
G_{{\mathcal{K}}_0}(\eta)G_{\{{\mathcal{K}}\}}(\eta)
=\sum_{n=0}^4f_n(\eta)X(n;\eta),
\end{equation}
where weight factors $f_n(\eta)$ are found to be
$f_0=f_4={\eta}^2, f_1=f_3=\sqrt{\eta}, f_2=1$,
and the operators $X(n;\eta)$ discriminate the basic
cluster charge configurations in accord with the central cluster
occupancy $n$:
\begin{eqnarray}
X(0;
\eta)&=&\prod_{k=1}^4P_k(0)Q_{\overline{k}}(\eta),\nonumber\\
X(1;
\eta)&=&\sum_{k=1}^4P_k(1)R_{\overline{k}}(\eta)\prod_{k'\neq
k}^4P_{k'}(0)Q_{\overline{k'}}(\eta),\label{a31}\\
X(2;
\eta)&=&\sum_{k<k'}P_k(1)P_{k'}(1)R_{\overline{k}}(\eta)
R_{\overline{k'}}(\eta) \prod_{k''\neq
k,k'}P_{k''}(0)Q_{\overline{k''}}(\eta),\nonumber\\
X(3;
\eta)&=&\sum_{k=1}^4P_k(0)Q_{\overline{k}}(\eta)\prod_{k'\neq
k}^4P_{k'}(1)R_{\overline{k'}}(\eta),\nonumber\\
X(4;
\eta)&=&\prod_{k=1}^4P_k(1)R_{\overline{k}}(\eta).\nonumber
\end{eqnarray}

The factorization procedure (\ref{a22}) constitutes the first step
of our approximation for calculating the expectation energy value
$E(\eta)$. As mentioned before, the second approximation consists
in the neglect of electron
spatial correlations within the Gutzwiller projected Fermi sea
state $|\psi_0\rangle=|\psi\rangle_{V=0}$, Eq.(\ref{a10}).
 For instance, when having a two-site
electron density correlation function
$\langle\psi_0|n_{k,\sigma}n_{k',\sigma'}|\psi_0\rangle$, we
decouple it as follows $(k\neq k')$:
\begin{equation}\label{a32}
\langle\psi_0|n_{k,\sigma}n_{k',\sigma'}|\psi_0\rangle
\simeq\langle\psi_0|n_{k,\sigma}|\psi_0\rangle
\langle\psi_0|n_{k',\sigma'}|\psi_0\rangle=\left(\frac{1}{2}n\right)^2
\end{equation}
where $n=N/L $ is an electron concentration and the extra factor
1/2 in the right-hand side is due to spatially uncorrelated
electron spins. Since doubly occupied sites are forbidden,
$\langle\psi_0|n_{k,\uparrow}n_{k,\downarrow}|\psi_0\rangle=0$, in
terms of site projection operators (\ref{a3}), the decoupling
(\ref{a32}) reads $(k\neq k')$:
\begin{equation}\label{a32bis}
\langle\psi_0|P_k(n)P_{k'}(n')|\psi_0\rangle \simeq\langle P_k(n)
\rangle_0 \langle P_{k'}(n')\rangle_0,
\end{equation}
where $\langle P_{k}(n)\rangle_0 = \langle\psi_0|P_{k}(n)|\psi_0\rangle=
1/2(\delta_{n,0}+\delta_{n,1})$. Multi-site density correlation
functions are decoupled in the same manner.

To complete this Section, let us calculate the norm
\begin{equation}\label{a33}
\langle\psi_{bc}(\eta)|\psi_{bc}(\eta)\rangle =\sum_{n=0}^4
f^2_n(\eta)\langle X(n; \eta^2)\rangle_0,
\end{equation}
where the average $\langle...\rangle_0$ is over the state $|\psi_0\rangle$.

By using the approximation (\ref{a32}) and (\ref{a32bis}) for a given
electron concentration $n=1/2$, one finds an intermediate result
\begin{equation}\label{a34}
\langle X(n; \eta^2)\rangle_0 =
\left(\frac{1}{2}\right)^4C^4_n\left(\langle
Q(\eta^2)\rangle_0\right)^{4-n}\left(\langle
R(\eta^2)\rangle_0\right)^n,
\end{equation}
where $C_n^4$ is a binomial and  both
$\langle Q_{\overline{k}}\rangle_0$ and $\langle
R_{\overline{k}}\rangle_0$ are calculated to be
$\overline{k}-$independent expectation values
$\langle Q(\eta^2)\rangle_0 =\langle R(\eta^2)\rangle_0
=({\eta}^4 + 4{\eta}+3)/8$.

From Eq.(\ref{a33}), the norm can be now written as
\begin{equation}\label{a39}
\langle
\psi_{bc}|\psi_{bc}\rangle = \left[\frac{{\eta}^4 + 4{\eta}+3}{8}\right]^5.
\end{equation}

\section{Calculation of the variational energy}
We start with the evaluation of an expectation value of the
Coulomb energy. Since the
procedure is similar to that used in the previous Section,
we present the resulting expression for the mean interaction energy
per cluster
\begin{eqnarray}
\frac{\langle\psi_{bc}|V_{{\mathcal{K}}_0}|\psi_{bc}\rangle}{\langle\psi_{bc}|\psi_{bc}\rangle}&=&
3V\frac{{\eta}^4+2{\eta}+1}{{\eta}^4 +4{\eta}+3}=\left\{
  \begin{array}{l}
    \frac{3}{2}V, \quad \eta=1 \\[3mm]
    V, \quad \eta=0
  \end{array}
\right.\label{a45}
\end{eqnarray}
Both limiting values in (\ref{a45}) can be easily understood on a
general ground. Indeed, the value $(3/2)V$ corresponds to spatially
uncorrelated electrons homogeneously distributed over the lattice,
while the value $V$ is in compliance with any lattice charge
configuration having \textit{two} electrons within each elementary
cluster.

The evaluation of the kinetic energy is much more involved. As a starting
point, one has to rely on a general expression
\begin{eqnarray}
&&\langle\psi_{bc}|T_{{\mathcal{K}}_0}|\psi_{bc}\rangle =-t\sum_{k\neq
k'}\sum_\sigma\langle
c^\dagger_{k\sigma}c_{k'\sigma}\rangle_{bc}=\nonumber\\
&&=-t\sum_{k\neq k'}\sum_{\sigma}\sum_{n=1}^3 f^2_n\langle X(n;
)c^\dagger_{k\sigma}c_{k'\sigma}X(n; )\rangle_0, \label{a46}
\end{eqnarray}
where each pair of sites $(k,k')\in{\mathcal{K}}_0$ contributes
equally to the sum (\ref{a46}); from now on we also use shortened
notation $X(n; \eta)=X(n; )$. To make the calculation more
transparent, we introduce projected fermionic operators
$\tilde{c}^\dagger_{k\sigma}(\tilde{c}_{k\sigma})$ related to
$c^\dagger_{k\sigma}(c_{k\sigma})$ as follows
\begin{equation}\label{a47}
\tilde{c}^\dagger_{k\sigma}=P_k(1)c^\dagger_{k\sigma}P_k(0),
\qquad \tilde{c}_{k\sigma}=P_k(0)c_{k\sigma}P_k(1).
\end{equation}
When operating in a space with no double site occupancy,
$\tilde{c}^\dagger_{k\sigma}(\tilde{c}_{k\sigma})$ and
$c^\dagger_{k\sigma}(c_{k\sigma})$ are equivalent.

Let us consider a particular pair $(k=1, k'=2)$ and insert into
(\ref{a46}) the explicit form of $X(n; )$ coming from (\ref{a31}).
 Then we obtain successively for
$n=1,2,3$ the following intermediate form of  hopping matrix
elements
\begin{eqnarray}
&&\langle X(1; )c^\dagger_{1\sigma}c_{2\sigma}X(1;
)\rangle_0=\nonumber\\
&&=\langle R_{\bar{1}}\tilde{c}^\dagger_{1\sigma}Q_{\bar{1}}\cdot
Q_{\bar{2}}\tilde{c}_{2\sigma}R_{\bar{2}}\cdot
P_3(0)Q^2_{\bar{3}}\cdot P_4(0)Q^2_{\bar{4}}\rangle_0,\nonumber\\
&&\langle X(2; )c^\dagger_{1\sigma}c_{2\sigma}X(2;
)\rangle_0=\label{a48}\\
&&=\langle R_{\bar{1}}\tilde{c}^\dagger_{1\sigma}Q_{\bar{1}}\cdot
Q_{\bar{2}}\tilde{c}_{2\sigma}R_{\bar{2}}\cdot\left[
P_3(0)Q^2_{\bar{3}}\cdot P_4(1)R^2_{\bar{4}}+
P_3(1)R^2_{\bar{3}}\cdot
P_4(0)Q^2_{\bar{4}}\right]\rangle_0,\nonumber\\
&&\langle X(3; )c^\dagger_{1\sigma}c_{2\sigma}X(3;
)\rangle_0=\nonumber\\
&&=\langle R_{\bar{1}}\tilde{c}^\dagger_{1\sigma}Q_{\bar{1}}\cdot
Q_{\bar{2}}\tilde{c}_{2\sigma}R_{\bar{2}}\cdot
P_3(1)R^2_{\bar{3}}\cdot P_4(1)R^2_{\bar{4}}\rangle_0.\nonumber
\end{eqnarray}
For brevity, in the above expressions we drop the arguments in
$Q_{\bar{k}}(\eta)$ and $R_{\bar{k}}(\eta)$.

It is worth discussing shortly the physical content of expressions
(\ref{a48}). An electron hopping within the central cluster is
affected by its charge surrounding, which may occur in many
configurations. In our description, all allowed configurations of
the charge surrounding are comprised by a set of operators
$Q_{\bar{k}}$ and $R_{\bar{k}}$ involved in the hopping matrix
elements (\ref{a48}). This becomes more clear if one uses the
explicit form of $Q_{\bar{k}}$ and $R_{\bar{k}}$, Eq.(\ref{a27}),
 which breaks up a hopping amplitude from
(\ref{a48}) into a sum of terms, each of them corresponds to a
particular configuration weighted with some $\eta-$dependent  factor.
In the uncorrelated limit, $\eta=1$, all these factors are unity,
while for $\eta<1$ hopping amplitudes are suppressed due to
short-range charge correlations.

The neglect of spatial electron correlations in $|\psi\rangle_0$
allows us to approximate the amplitudes (\ref{a48}) by using a
decoupling procedure. For instance, the first expression in
(\ref{a48}) is decoupled  as follows
\begin{eqnarray}
&&\langle X(1; )c^\dagger_{1\sigma}c_{2\sigma}X(1;
)\rangle_0\simeq\nonumber\\
&\simeq&\langle R_{\bar{1}}Q_{\bar{1}}\rangle_0\langle
R_{\bar{2}}Q_{\bar{2}}\rangle_0\langle Q^2_{\bar{3}}\rangle_0
\langle Q^2_{\bar{4}}\rangle_0\langle P_3(0)\rangle_0\langle
P_4(0)\rangle_0\langle
\tilde{c}^\dagger_{1\sigma}\tilde{c}_{2\sigma}\rangle_0=\nonumber\\
&=&\left(1/2\right)^2\langle RQ\rangle^2_0\langle
Q^2\rangle^2_0\langle
c^\dagger_{1\sigma}c_{2\sigma}\rangle_0,\label{a49}
\end{eqnarray}
and the remaining two amplitudes in (\ref{a48}) are approximated in
the same way.
In (\ref{a49}) the last equality takes into account
$\bar{k}-$independence of expectation values $\langle
R_{\bar{k}}Q_{\bar{k}}\rangle_0 = \langle RQ\rangle_0,\quad
\langle Q^2_{\bar{k}}\rangle_0=\langle Q^2\rangle_0$ and the
property $\langle
\tilde{c}^\dagger_{1\sigma}\tilde{c}_{2\sigma}\rangle_0 = \langle
c^\dagger_{1\sigma}c_{2\sigma}\rangle_0$.
The expression for
$\langle Q^2\left(\eta\right)\rangle_0=\langle
Q\left(\eta^2\right)\rangle_0,  \langle
R^2\left(\eta\right)\rangle_0=\langle
R\left(\eta^2\right)\rangle_0$,
is given in the previous Section and the
expectation value $\langle R_{\bar{k}}\left(\eta\right)
Q_{\bar{k}}\left(\eta\right)\rangle_0=\langle RQ\rangle_0$
is calculated easily to give
$\langle RQ\rangle_0=\sqrt{\eta }\left({\eta}^2+3\right)/4$.

Next, by collecting these results and with
the reference to (\ref{a46}), we obtain the following
relation between hopping amplitudes in the
correlated, $V\neq0$, and 'uncorrelated', $V=0$, states
\begin{equation}\label{a53}
\langle c^\dagger_{k\sigma}c_{k'\sigma}\rangle_{bc}=\frac{2}{8^4}
{\eta}
\left(\eta+1\right)\left({\eta}^2+3\right)^2\left({\eta}^4+4\eta+3\right)^2
\langle c^\dagger_{k\sigma}c_{k'\sigma}\rangle_0,
\end{equation}
 The latter hopping amplitude $\langle
c^\dagger_{k\sigma}c_{k'\sigma}\rangle_0$ is obviously related to
the expectation value of the kinetic energy per cluster calculated
at $V=0$:
\begin{equation}\label{a54}
\langle T_{{\mathcal{K}}_0}\rangle_0 = -t\sum_{\stackrel{k\neq
k'}{(k,k')\in {{\mathcal{K}}_0}}}\langle
c^\dagger_{k\sigma}c_{k'\sigma}\rangle_0.
\end{equation}
This leads to the final result
\begin{equation}\label{a55}
\frac{\langle\psi|T_{{\mathcal{K}}_0}|\psi\rangle_{bc}}{\langle\psi|\psi\rangle_{bc}}
=q(\eta)\langle T_{{\mathcal{K}}_0}\rangle_0,\quad
q(\eta)=\frac{16\eta\left(\eta+1\right)\left({\eta}^2+3\right)^2}
{\left({\eta}^4+4\eta+3\right)^3},
\end{equation}
where $q(\eta)$ is the kinetic energy renormalization factor. Note
that in the 'uncorrelated' limit, $V=0$, the necessary condition
$q(\eta=1)=1$ is fulfilled.
The kinetic energy (\ref{a55}) and the Coulomb term (\ref{a45})
together give the final result for the variational energy $E(\eta)$
per cluster.

In the extreme limit, $V/t\to\infty$, all clusters are doubly
occupied (the 'tetrahedron rule' is fulfilled completely), which
leads to the upper bound of the variational energy, $E_{max}=V$,
at $\eta=0$. Let us consider
\begin{equation}\label{a57}
E(\eta)-E(\eta=0)= \frac{2\eta \left({\eta}^3 + 1 \right
)}{\left({\eta}^4 + 4 \eta + 3 \right)}\left[V -
8|\bar{\mathcal{E}}|\frac{\left({\eta}^2+3\right)^2}
{\left({\eta}^2 -\eta+1\right)\left({\eta}^4+4\eta+3\right)^2}
\right]\leq 0,
\end{equation}
where $\langle
T_{{\mathcal{K}}_0}\rangle_0\equiv-|\bar{\mathcal{E}}|<0$.
The expression between the brackets in (\ref{a57}) changes sign
at some critical coupling $V=V_{c}$, with the following requirement
\begin{equation}\label{a58}
\eta=\left\{
  \begin{array}{l}
    0, \quad V>V_{c};\\
    >0, \quad V<V_{c};
  \end{array}
\right.
\end{equation}
A close inspection of Eq. (\ref{a57})  suggests the critical
value $V_{c}=8|\bar{\mathcal{E}}|$. To
check this suggestion, let us minimize $E(\eta)$ by solving the
equation  $\partial E(\eta)/\partial\eta=0$, which relates $\eta$ to
$V$ and requires the only solution
$\eta=0$ at ${V}={V}_{c} = 8|\bar{\mathcal{E}}|$.
In the vicinity of ${V}_{c}$, i.e. for ${V}\lesssim
 8|\bar{\mathcal{E}}| $,  the kinetic energy renormalization
factor is given by
\begin{equation}\label{a61}
q \simeq \frac {3}{2}\left(1-\frac{V}{V_c}\right),
\end{equation}
which is checked numerically to describe the mean kinetic energy
reduction with high accuracy in a  wide range of coupling, $1/2
< V/V_c \leq 1$.
Rather generally, one may estimate the value of $\langle
T_{{\mathcal{K}}_0}\rangle_0$ to  be  $\bar{\mathcal{E}} = -azt$,
where the coefficient $a$ is of order  unity. Therefore, according
to our convention, the solution (\ref{a61}) applies to the strong
coupling regime of the model (\ref{a1}), where the wave function
ansatz in the form (\ref{a18}) is  valid.  We do  not intend
to discuss here the variational ground state energy in the  weak coupling
limit, $V \lesssim t$.  As argued in
Sections II, in this limit
kinetic energy effects dominate inter-site Coulomb
correlations, the wave function ansatz in the form (\ref{a18}) is
not applicable any more and its complementary variational
extension is required.

 The physical picture emerging from these results is discussed
in the following Section. The discussion is closely related to
 results of the Brinkman-Rice  theory. \cite{Brinkman70}


\section{Discussion.}
 Based on a simple trial wave function (\ref{a18}) with a single
 variational parameter $\eta$, we infer from the results (\ref{a55}),
 (\ref{a57}) and (\ref{a61}) that in the strong coupling regime, $V\gg
 t$, the mean kinetic energy per cluster, $q \langle
 T_{{\mathcal{K}}_0}\rangle_0$, of the quarter-filled pyrochlore
 electronic system is strongly reduced by the factor $q$ ($\rightarrow 0$, as
 $V\rightarrow V_{c}-0 $). This reduction means the increasing
 electron localization in real space and thus a strongly correlated
 metallic state is expected to appear for $V\lesssim V_c$. For $V>
 V_c$, the mean kinetic energy is zero and the variational energy is
 $V$ per cluster, which describes an insulating highly degenerate state
 of fully
 localized electrons. Such a description of the insulating state is
 correct, however, only in the limit of extreme coupling,  $V/t\to\infty$.
 For finite, however, still large $V/t$,  both in the pyrochlore and
 checkerboard lattices\cite{Runge04}  charge fluctuations to leading order in
 $t/V$ would lower the ground state energy by the value of order $t^2/V$.
 It means that even in an insulating state, full
 electron localization cannot be reached. This observation casts doubt
 on the very existence of the predicted localization transition.

 In this respect, our results are very
 similar to those obtained by Brinkman and Rice in their  study
 \cite{Brinkman70} of the  Hubbard model based on
 Gutzwiller's wave function and approximation in the limit of a
 half-filled band. There, the driving force for electron
 correlations and localization, at a finite  strength
 $U_c$,  is the strong on-site Hubbard interaction $U$, while in the present
 study  of the quarter band filling the electron
 localization is due to  inter-site Coulomb interaction $V$.
The merits and failures of Brinkman-Rice  theory are well
understood and widely discussed in the literature (see Refs.
\onlinecite{Vollhardt84,Georges96,Imada98,Gebhard97}, and references
therein).
It was realized, for
instance, that the simple Gutzwiller wave function\cite{Gutzwiller65}
is not rich enough to describe the true insulating state in the
Hubbard model and that at any finite dimension of a lattice
the occurrence
of the Brinkman-Rice localization transition is merely due to the
Gutzwiller approximation used. According to this approximation, the
spatial correlations in the system of noninteracting electrons  are
neglected. (In the present study we neglect spatial correlations in the
Gutzwiller projected Fermi sea (\ref{a10}) which is the
starting wave function chosen before applying a generalized Gutzwiller
projector (\ref{a43})). This finding, however,  does not
invalidate the general conclusion about the usefulness of the
Gutzwiller approximation which is known to give a proper physical
description in a number of situations. Indeed, as shown by Vollhardt,
W\"olfle and Anderson\cite{VWA87}, a Hubbard lattice-gas model
in the not yet localized regime ($U/U_c \lesssim 1$) 
describes ground state properties of normal liquid $^3$He
('almost localized' Fermi liquid \cite{Vollhardt84}).

In view of the above discussion, our results suggest that in the
quarter-filled pyrochlore lattice a metallic state of nearly localized
electrons may occur mainly due to strong enough intersite Coulomb
interaction, i.e. for $V/V_c\lesssim 1$. To our knowledge, the present
variational theory is the
first one showing the importance of short-range Coulomb interaction in
forming a strongly correlated metallic state in the pyrochlore
lattice. Calculation of physical properties of such a state is,
however, beyond the scope of the present study. Therefore, many
questions remain to be answered.  For instance, whether the predicted
state is the Fermi liquid and, if it is so, may one relate the
renormalization parameter $q$ to  the discontinuity at the Fermi
vector of the momentum distribution function? An answer would allow
us to connect the inverse of $q$ to a charge quasiparticle mass
enhancement  in the
strongly correlated regime of the model. Concerning  a description of
 the weak
interaction case, $V\lesssim t$, the trial wave function
(\ref{a18}) must be extended, for instance, by introducing a
second variational parameter, which would make the present theory
more attractive. We believe, however, such an extension of
(\ref{a18})  does not change our main results, and retain these
problems for a further study.

\vspace*{1cm}

{\bf Acknowledgments}

The authors are grateful to K. Kikoin for useful discussions.

\newpage
\begin{figure}
\includegraphics[width=7cm]{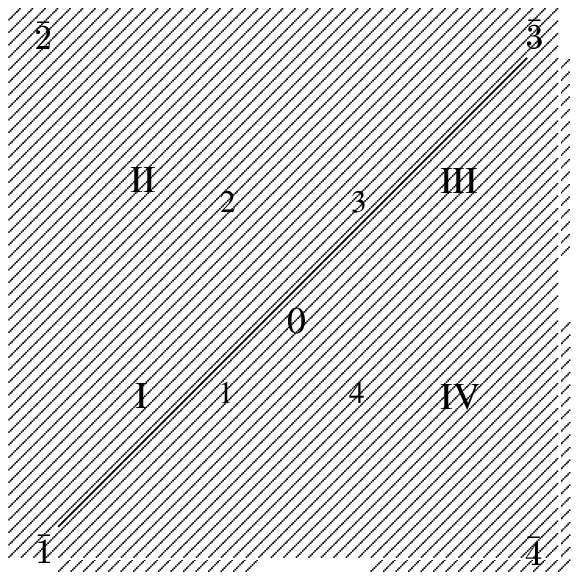}
\caption{Fragment of the checkerboard lattice chosen to be a
minimal basic cluster in the variational  cluster  calculations.
Each label ${\mathcal{K}}$  of side clusters
$(\mathrm{I},\ldots,\mathrm{IV})$ is a composite
one, ${\mathcal{K}}=(k,\overline{k})$, where $k$ ($=1,...,4$) is
referred to a site  shared with the central 0-cluster and
$\overline{k}$ ($=\overline{1},...,\overline{4}$)  is for the
complementary 3-site fragment (hatched). In the pyrochlore
lattice, the same labelling is used  for a minimal basic cluster
formed by five corner-sharing tetrahedra.}
\end{figure}

\end{document}